\documentclass[12pt,aps,notitlepage,superscriptaddress, amsmath,amssymb,amsfonts,floatfix]{revtex4}
\usepackage{amsmath}
\usepackage{graphicx}
\usepackage{epsfig}
\usepackage{amssymb}

\usepackage{color}

\newcommand{\be}{\begin{equation}}
\newcommand{\ee}{\end{equation}}
\newcommand{\bea}{\begin{eqnarray}}
\newcommand{\eea}{\end{eqnarray}}

\begin{document}

\title{Collective canard explosions of globally-coupled rotators with adaptive coupling}

\author{Marzena Ciszak}
	\affiliation{CNR - Consiglio Nazionale delle Ricerche - Istituto Nazionale di Ottica, Via Sansone 1, I-50019 Sesto Fiorentino (FI), Italy}
\author{Simona Olmi}
	\affiliation{CNR - Consiglio Nazionale delle Ricerche - Istituto dei Sistemi Complessi, via Madonna del Piano 10, I-50019 Sesto Fiorentino, Italy}	 
	\affiliation{INFN, Sezione di Firenze, Via Sansone 1, I-50019 Sesto Fiorentino (FI), Italy}	
	\author{Giacomo Innocenti}
	\affiliation{Department of Industrial Engineering, University of Florence, via di Santa Marta 3, 50139 Florence, Italy}
	\author{Alessandro Torcini}
	\affiliation{Laboratoire de Physique Th\'eorique et Mod\'elisation, Universit\'e de Cergy-Pontoise,CNRS, UMR 8089, 95302 Cergy-Pontoise cedex, France}
	\affiliation{CNR - Consiglio Nazionale delle Ricerche - Istituto dei Sistemi Complessi, via Madonna del Piano 10, I-50019 Sesto Fiorentino, Italy}
	\author{Francesco Marino}
	\affiliation{CNR - Consiglio Nazionale delle Ricerche - Istituto Nazionale di Ottica, Via Sansone 1, I-50019 Sesto Fiorentino (FI), Italy}
	\affiliation{INFN, Sezione di Firenze, Via Sansone 1, I-50019 Sesto Fiorentino (FI), Italy}	
	\date{\today}

\begin{abstract}
Canards, special trajectories that follow invariant repelling slow manifolds for long time intervals, have been frequently observed in slow-fast systems of either biological, chemical and physical nature. Here, collective canard explosions are demonstrated in a population of globally-coupled phase-rotators subject to adaptive coupling. In particular, we consider a bimodal Kuramoto model displaying coexistence of asynchronous and partially synchronized dynamics subject to a linear global feedback. A detailed geometric singular perturbation analysis of the associated mean-field model allows us to explain the emergence of collective canards in terms of the stability properties of the one-dimensional critical manifold, near which the slow macroscopic dynamics takes place. We finally show how collective canards and related manifolds gradually emerge in the globally-coupled system for increasing system sizes, in spite of the trivial dynamics of the uncoupled rotators.
\end{abstract}
 
\maketitle

\section{Introduction}
\label{intro}

Slow-fast systems, i.e. nonlinear dynamical systems in which two or more variables evolve over very different time-scales, appear in many branches of natural science. The splitting between time-scales gives rise to peculiar oscillatory patterns, the dynamics of which can be generally decomposed into a sequence of periods of slow motion, taking place near the attracting branches of a critical manifold defined by the equilibria of the fast dynamics, and rapid switches between them (relaxation and bursting oscillations) \cite{smale}. Remarkably, at the transition from small-amplitude (Hopf-type) limit cycles to large relaxation/burst cycles, the trajectories may flow on a slow time-scale along the repelling part of the manifold, instead of quickly departing from it. These special trajectories are known as {\it canard cycles} \cite{canard}. The transition (canard explosion) occurs within an exponentially small range of parameter values, which makes these orbits hard to detect numerically and observe experimentally. Moreover, since canard orbits partially connect attracting and repelling slow manifolds, they are extremely sensitive to variations of control parameters and fluctuations. Such sensitivity is also at the origin of complex dynamics including chaotic spiking and mixed-mode oscillations \cite{brons,krupa,Guck1,Guck2,desroches2012,desroches2016,marino2007,alnaimee2009,alnaimee2010,marino2011}.

While most studies on canard phenomena have been carried out in low-dimensional slow-fast systems, particularly two- or three-dimensional neuronal models \cite{izhikevich,desroches2013,mitry2006,moehlis,kramer2008}, recent investigations focused on large populations \cite{touboul,dolcemascolo,otti}. These works have demonstrated, e.g., the role of large networks in producing robust collective responses and preserving canard orbits in the presence of noise, which has important implications in neuroscience. Other studies more generally focused on collective slow-fast dynamics, such as bursting \cite{ivan,igor} and excitability \cite{marzena2009}. In all these cases however, these phenomena result in the partial synchronization of the nodes, each one being an autonomous slow-fast system.

In this work, we show how canard explosions can spontaneously emerge in the collective dynamics in large populations of globally-coupled phase oscillators.


Complex networks of rotators, typically described in terms of the Kuramoto model \cite{kuramoto} and its generalizations \cite{acebron2005}, have been widely investigated in the last decades with the aim of observing the emergence of non-trivial macroscopic phenomena. These includes e.g. collective oscillations, quasi-periodicity and chaos \cite{matthews1990, hakim1992, nakagawa1993}, and first-order phase transitions between states with different degrees of synchronization \cite{tanaka1997,pazo2009,olmi2014}. However, only a few pioneering analyses have been devoted to the onset of collective slow-fast phenomena in such networks \cite{so2011,skardal2014}. 

In a recent paper \cite{noi} some of us have shown that a self-sustained adaptation mechanism can give rise to collective excitability and bursting oscillations in a population of rotators. In the absence of adaptation, the network display a hysteretic phase transition involving asynchronous and partially synchronized states. The adaptive feedback drives the system on a slow time-scale across the phase-transition, leading to self-sustained collective oscillations that can be either periodic or chaotic. These dynamics are fairly reminiscent of the spiking and bursting dynamics observed in the Hindmarsh-Rose model for a single neuron \cite{hr,wang,gmiranda,innocenti2007}. However, many important features related to these phenomena, in particular the transitional regime from the Hopf quasi-harmonic cycle and the fully-developed bursting, have not been explored so far. 

Here, we focus on this regime showing the existence of collective canard explosions in a large population of phase oscillators subject to a linear adaptive coupling. We demonstrate that these phenomena originate from the existence of a one-dimensional (1D) critical manifold which organizes the mean-field dynamics on a slow time-scale. We perform extensive numerical simulations to characterize canard trajectories as a function of the control parameters and to address the effects of finite-size fluctuations.
 
The paper is organized as follows. In Sec. II, we introduce the network model and its mean-field description, consisting of a 3D slow-fast system with two fast and one slow variable. In Sec. III we present the collective dynamical regimes encountered in our network and compare them with the prediction of the mean-field model. In Sec. IV we focus on the fast transition from the small-amplitude Hopf cycle to the large amplitude bursting oscillations and explain it in terms of classical canard explosions nearby a 1D critical manifold, $\Sigma$. We finally show in Sec. V that canard cycles associated to effective slow-manifolds in the phase space, are already observable even at low values of $N$. Remarkably, such a geometric structure is not encoded in the single-node dynamics, but gradually emerge in the macroscopic behaviour of the network. Conclusions and future perspectives are presented in Sec. IV.

\section{Adaptive Kuramoto model and mean-field description}
\label{models}

We consider a globally-coupled population of $N$ rotators with adaptive coupling strength $S(t)$ \cite{noi}
\begin{subequations}
\label{network} 
\begin{eqnarray}
{\dot \theta}_k(t) &=& \omega_k + \frac{S(t)}{N} \sum_{j=1}^N \sin(\theta_j(t)- \theta_k(t)), \;  \qquad k=1,\ldots, N \label{net1} \\
{\dot S}(t) &=& \epsilon \left[-S(t) + K -\alpha R(t) \right] \;  \label{net2}
\end{eqnarray}
\end{subequations}
where $\theta_k$ ($\omega_k$) are the phases (natural frequencies) of each rotator. Accordingly to Eq. \eqref{net2}, the evolution of the coupling variable $S(t)$ is controlled, via a linear feedback, by $R(t)$, which is the modulus of the complex Kuramoto order parameter $Z(t) = \frac{1}{N} \sum_{j=1}^N {\rm e}^{i \theta_j(t)} = R(t) {\rm e}^{i \phi(t)}$ \cite{kuramoto}.
The macroscopic variable $R$ measures the level of synchronization among the rotators: asynchronous (partially synchronized) dynamics will correspond to $R = 0$ ($0 < R \le 1$). We consider a bimodal distribution of natural frequencies for which, in absence of feedback ($\alpha=0$), the system displays a first-order hysteretic transition from incoherent to partially-synchronized states \cite{pazo2009}. The gain of the feedback loop is controlled by $\alpha$ while $\epsilon$ is the ratio between the characteristic time-scale of the macroscopic network dynamics and the feedback. We assume the coupling to slowly adapt to the fast switching between incoherent and coherent states, i.e. $0 < \epsilon \ll 1$. It is in this regime that the multiple time-scale competition between the macroscopic network dynamics and the feedback gives rise to collective slow-fast dynamics \cite{noi}.

For the special case of a bimodal frequency distribution given by the sum of two Lorentzians an exact mean-field model for the network \eqref{network} can be obtained by extending the calculations in \cite{martens2009,so2011} based on the the Ott-Antonsen Ansatz \cite{ott2008}. 
Writing $Z = \frac{1}{2}(z_1 + z_2)$ in terms of two sub-population order parameters $z_k=\rho_k{\rm e}^{i \phi_k}$ ($k=1,2$) each relative to a Lorentzian distribution, and assuming $\rho_1 \approx \rho_2 = \rho$ we obtain \cite{noi}
\begin{subequations}
\label{mf} 
\begin{eqnarray}
\dot{\rho} & = & -\Delta \rho + \frac{S}{4} \rho (1 - \rho^2) (1 + \cos(\phi)) \;  \label{eq:model1} \\
\dot{\phi} & = & 2 \omega_0 - \frac{S}{2} (1 + \rho^2) \sin(\phi) \;  \label{eq:model2} \\
\dot{S} & = & -\epsilon \left[ S - K + \alpha \rho \sqrt{\frac{1 + \cos(\phi)}{2}} \right] \;  \label{eq:model3} 
\end{eqnarray}
\end{subequations}
where $\phi = \phi_2 - \phi_1$, $\pm \omega_0$ are the centers of the two Lorentzians, $\Delta$ is their half-width at half-maximum and $R = \rho \sqrt{(1 + \cos(\phi))/2}$. In this paper, we keep fixed the distribution parameters at the values $\omega_0=1.8$ and $\Delta=1.4$ and the feedback gain at $\alpha=7$.

The stability of Eqs. (\ref{mf}) can be analyzed by following the evolution of infinitesimal perturbations in the tangent space, whose dynamics is ruled by the linearization of Eqs. (\ref{mf})  as follows:
\begin{eqnarray}
\label{mf_tangent}
\nonumber
\delta\dot{\rho} & = & -\Delta \delta\rho + \left[ \frac{S}{4} (1 + \cos(\phi)) -\frac{S}{4} \sin(\phi)\delta\phi \right] \rho (1 - \rho^2) + \frac{S}{4} (1 + \cos(\phi))(1-3\rho^2) \delta\rho\; \\
\delta\dot{\phi} & = & 2 - \frac{\delta S}{2} (1 + \rho^2) \sin(\phi) - \frac{S}{2} (1 + \rho^2) \cos(\phi)\delta\phi -S\rho\sin(\phi) \delta\rho\;  \\ \nonumber
\delta\dot{S} & = & -\epsilon \left[ \delta S + \alpha \sqrt{\frac{1 + \cos(\phi)}{2}} \delta\rho -\frac{\alpha\rho}{8} \frac{\sin(\phi)}{\sqrt{1+\cos(\phi)}} \delta\phi\right]. \;  
\end{eqnarray}
In particular the maximal Lyapunov exponent $\lambda_M$ associated to the mean-field model \eqref{mf} can be obtained by considering the time evolution of the tangent vector $\vec{\delta}$= $\left\lbrace \delta\rho, \delta\phi, \delta S \right\rbrace$, ruled by the linearized version of the original system \eqref{mf_tangent}. The maximal Lyapunov quantifies the average growth rate of an infinitesimal perturbation $\vec{\delta}_0$ and it can be estimated as follows
\begin{equation}
 \lambda_M= \lim_{t\rightarrow\infty}\frac{1}{t}\log\frac{|\vec{\delta}(t)|}{|\vec{\delta}_0|};
\end{equation}
the system is chaotic whenever $\lambda_M>0$ \cite{pikovsky2016}. As we shall see in Sec. \ref{canards}, our system display a regime of irregular canard cycles the characterization of which in terms of the asymptotic maximal LE is extremely difficult. For such states it is more useful to estimate the finite time Lyapunov exponent $\lambda^t$ over a finite time window of duration $\Delta t$, namely $\lambda^t=\frac{1}{\Delta t}\ln\sqrt{\sum_{i=1}^3 \delta_i(\Delta t)\delta_i(\Delta t)}$, where the initial magnitude of the vector is set to one, i.e. $||\vec{\delta}(0)||\equiv 1$. 

\section{Network and mean-field dynamics}
\label{results1}
 
We now present the dynamical regimes obtained integrating the network model (\ref{network}) and compare them with the corresponding mean-field predictions. Unless noted otherwise, in our study the bimodal Lorentzian distributions are randomly generated through the rule $\omega_j = \pm \omega_0 +\Delta \tan(\pi(r - 0.5))$, where $r$ is a random number uniformly distributed between 0 and 1. Alternatively, we can generate them deterministically, as follows \cite{montbrio2015}
\begin{subequations}
\label{gend} 
\begin{eqnarray}
\omega_j &=& -\omega_0 + \Delta \tan (\pi\xi_j/2)  \qquad \xi_j = \frac{2j-N/2-1}{N/2+1} \qquad j=1 \dots N/2\\
\omega_j &=& \omega_0 + \Delta \tan (\pi\xi_j/2) \qquad \xi_j = \frac{2j-3N/2-1}{N/2+1} \qquad j=N/2+1 \dots N
\end{eqnarray}
\end{subequations}  
While for large $N$ ($N\gtrsim 10^5$) our results are not modified qualitatively by the way in which the distribution is generated (nor by considering other bimodal frequency distributions, such as Gaussian), we will show in Sec. \ref{emergent} that this is not the case at lower values of $N$. 

\begin{figure}
\begin{center}
\includegraphics[width=0.7\textwidth]{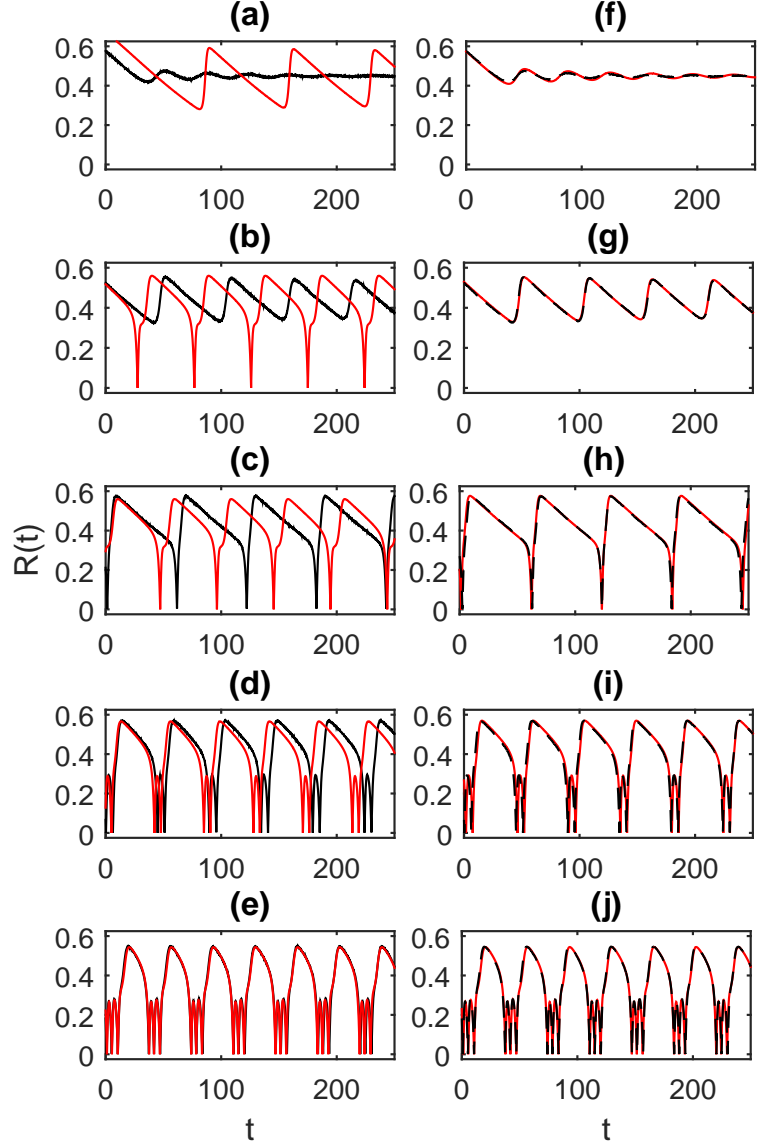}
\end{center}
\caption{Time-series of $R$ for the network model (\ref{network}) (black) and the mean-field model (\ref{mf}) (red). In each panel from (a) to (e) the control parameter $K$ used in both models is the same: (a) $K=$9.044, (b) $K=$9.017137, (c) $K=$9.017, (d) $K=$8.9, (e) $K=$8.5. In panels (f)-(j) we plot the same network time-traces as in (a)-(e), while the control parameter in the mean-field model has been adjusted to reproduce the dynamics of the network: (f) $K=$9.06, (g) $K=$9.04441, (h) $K=$9.044288, (i) $K=$8.925, (j) $K=$8.505. Other parameters: $\epsilon=$0.02, $\Delta_0$=1.4 
$\omega_0$=1.8, $\alpha$=7. For the network we use a population of $N=5 \times 10^5$ rotators and a randomly-generated bimodal Lorentzian distribution.}
\label{figure1}
\end{figure}

Time-traces of the different macroscopic regimes illustrated in terms of the synchronization parameter $R(t)$ are displayed in Fig. \ref{figure1}. For $K=$9.044 the network is in a partially-synchronized state, associated with a constant synchronization variable $R(t) \approx 0.45$. A supercritical Hopf bifurcation occurs at the critical value $K=9.017135$ beyond which $R(t)$ displays quasi-harmonic small-amplitude oscillations (see e.g. Fig. \ref{figure1} (b)). This behavior is observable only within a parameter range of order $\epsilon$ beyond the bifurcation point. Outside this range the system enters the bursting regime, alternating partially synchronized phases with abrupt de-synchronization events (spikes), as shown in Fig. \ref{figure1} (c). Further decreasing $K$, the number of spikes increases in a spike-adding sequence (see $2$-spike and $3$-spike bursting in (Fig. \ref{figure1} (d,e)) until the number of spikes becomes irregular, giving rise to a chaotic bursting phase. Such macroscopic dynamics is quite similar to that observed in low-dimensional slow-fast systems, such as the Hindmarsh-Rose model \cite{innocenti2007} and has been investigated in detail in Ref. \cite{noi}.

In this work we focus instead on the fast transition from the quasi-harmonic Hopf cycle (Fig. \ref{figure1} (b)) to the single-spike bursting regime (Fig. \ref{figure1} (c)). In a small range of $K$ between these regimes, the amplitude of the limit cycle abruptly (though continuously) grows and reaches a saturation value. Likewise, the frequency of the oscillations experiences a similar sudden change from the quasi-harmonic value to a frequency of the order of $\epsilon$ typical of the bursting regime. 
Close to the transition both the Hopf and burst cycles are extremely sensitive to variations of the control parameter and to fluctuations. As a result, the mean-field time-series quantitatively differ from those of the network, in spite of the large number $N$ of oscillators involved. In Fig. \ref{figure1} (b), the network and the mean-field model are neither showing the same dynamical regime. A better agreement is found far from the transition, where the system displays multi-spike bursting oscillations (see Fig. \ref{figure1}(c-d)). These attractors typically exists over a wider range of parameters and are less sensitive to fluctuations or variations of $K$.

On the other hand, it is sufficient to slightly tune the control parameter $K$ towards lower values, to find an excellent agreement between the macroscopic dynamics of the network and the mean-field predictions. This suggests that the finite size of the network has the main effect (at least at the level of $N \sim 10^5$ oscillators) of shifting the bifurcation points, while having a moderate effect on the orbits in phase space. In the case of a deterministically-generated frequency distribution, the network dynamics is generally better approximated by the mean-field. We will return back to this point in Sec. V. 

While the above scenario displays all the characteristic features of canard explosions, a rigorous interpretation in terms of the classical canard phenomenon requires the identification of the associated geometric structures from which slow segments of trajectories are attracted and repelled. This is what we discuss in the next section.

\section{Collective canard explosions}
\label{canards}

In geometric terms a (maximal) canard solution originates from the connection of attracting and repelling slow manifolds near non-hyperbolic points that, for $\epsilon \ll 1$, are approximated by the critical manifold defined by the set of equilibria of the fast dynamics \cite{fenichel1979}. Other canard solutions exist in an exponentially small neighbourhood of maximal canards. As we shall see below, in the appropriate parameters range a periodic pattern of canard cycles may appear as the result of slow motion nearby the critical manifold, combined with the turning back of the fast dynamics.

On the fast time scale $t$, the evolution is described by the mean-field equations (\ref{eq:model1}-\ref{eq:model2}) (layer problem) with $S$ acting as a bifurcation parameter. The equilibria of this dynamical subsystem lay on the one-dimensional manifold $\Sigma= \Sigma_0 \cup \Sigma_{\rho}$, where $\Sigma_0$ is given by the set of incoherent steady-state solutions $\Sigma_0 = \{\rho_s = 0, \sin(\phi_s) = 4 \omega_0 /S, S\}$, and $\Sigma_{\rho}= \{\rho_s, \phi_s, S\}$ is defined by the equations $S (1+\rho_s^2) \sin(\phi_s)=4 \omega_0$ and 
\begin{equation}
S =\frac{2\omega_0^2}{\Delta} \frac{1-\rho^2}{(1+\rho^2)^2} + \frac{2 \Delta }{1-\rho^2} \equiv \mathcal{F}(\rho)
\label{eq_cm}
\end{equation}
On the slow time scale $\tau=\epsilon t$, the motion is governed by the feedback equation Eq. (\ref{eq:model3}) (reduced problem) with the algebraic constraint $(\dot{\rho},\dot{\phi})$=($0,0$). The fixed points of the layer problem thus define the 1D critical manifold near which the slow dynamics takes place. In particular, the trajectories will be attracted by stable parts of $\Sigma$ while will be repelled by the unstable ones and the dynamics will be well approximated by that of the reduced problem \cite{fenichel1979}.

\begin{figure}
\begin{center}
\includegraphics[width=0.7\textwidth]{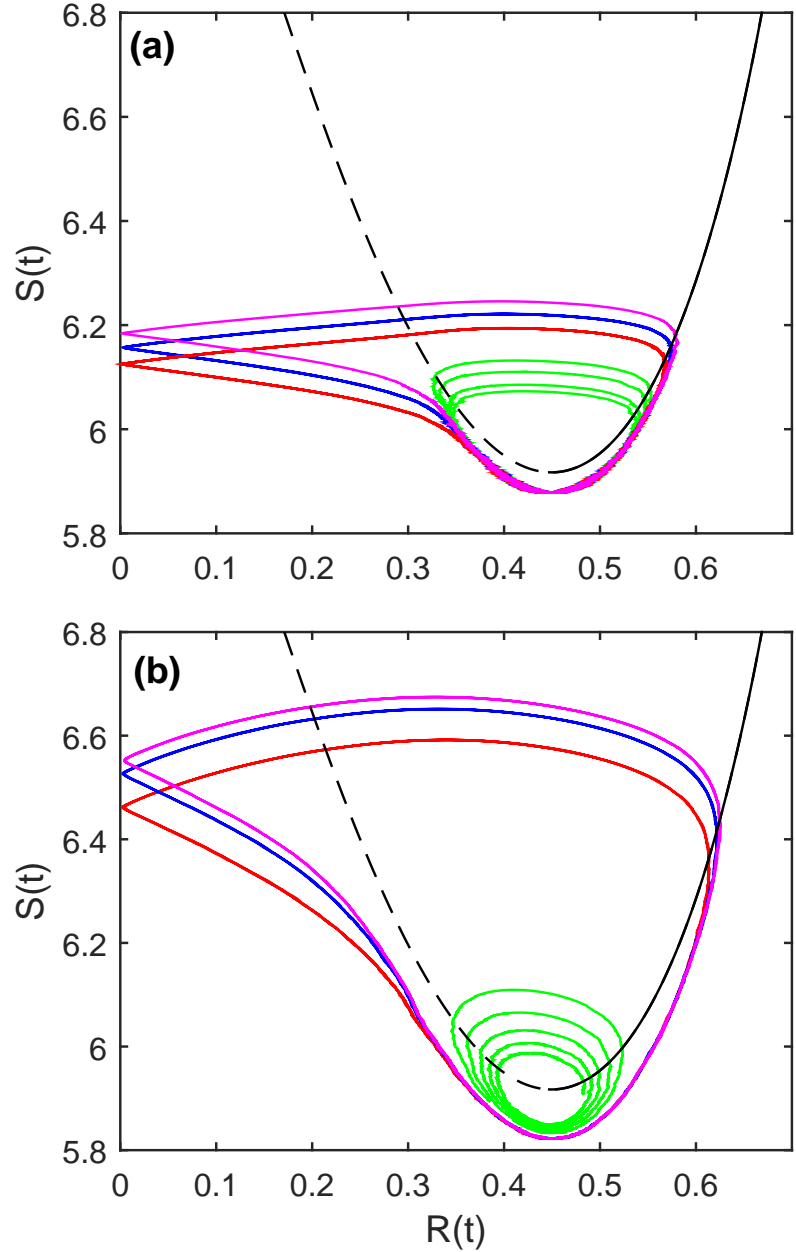}
\end{center}
\caption{Hopf and canard cycles of the network model (\ref{network}) for (a) $\epsilon=$0.02 and (b) $\epsilon=$0.07. The limit cycles are illustrated in the ($S$,$R$) plane together with the critical manifold (\ref{smr}, whose attracting (repelling) branches are plotted as black solid (dashed) lines. 
(a) Hopf cycle $K=$9.017137 (green) and canard cycles $K=$9.0165 (red), $K=$9.017 (blue), $K=$9.0171 (magenta). (b) Hopf cycle  $K=8.98$ (green) and canard cycles $K=$8.97 (red), $K=$8.9725 (blue), $K=$8.973. Other parameters as in Fig. \ref{figure1}.}
\label{fig2}
\end{figure}

Linearizing the fast subsystem on $\Sigma_{\rho}$ we find that it consists of a branch of stable equilibria $\Sigma_S$ (see e.g. solid black line in Fig. \ref{fig2}) and an unstable one $\Sigma_R$ (dashed line) coalescing in a saddle-node bifurcation at a fold point.
For $\omega_0 > \Delta$, equilibria along $\Sigma_0$ are always unstable. Finally, the above stationary states coexist with a multiplicity of stable limit cycles which are responsible for the onset of the bursting phase and the spike adding-sequences of the complete system \cite{noi}.

In order to verify the existence of canard trajectories in our network, we re-write the critical manifold (\ref{eq_cm}) in terms of $R$, using the relation $R = \rho \sqrt{(1 + \cos(\phi))/2}$ and the stationary solutions of Eq. \ref{eq:model1}. We obtain the implicit expression
\begin{equation}
S =(S R^2 + 2 \Delta)\left(1+ \frac{\omega_0^2}{S R^2 + \Delta} \right)
\label{smr}
\end{equation}

In Fig. \ref{fig2}(a) we plot the projection on the $(R, S)$ plane of a few solutions of Eqs. (\ref{eq:model1}-\ref{eq:model3}) as $K$ is varied, together with the critical manifold (\ref{smr}). As often observed in planar slow-fast systems, canard cycles, i.e. phase-space orbits with portions closely following the repelling branch of the critical manifold, arise in the vicinity of the fold point where normal hyperbolicity is lost via a saddle-node bifurcation of the layer problem. The trajectories passes close to the repelling branch $\Sigma_R$ for a certain amount of time, that critically depends on the control parameter $K$. For higher values of $\epsilon$, the slow dynamics is less constrained by the critical manifold and the slow-fast character of the orbits is less pronounced. In particular, both canards and Hopf cycles tend to become more circular (see \ref{fig2}(b)), and for very large $\epsilon$ (of a few tenths) canard cycles no longer exist, becoming fully convex and loosing their characteristic inflection point. 

On the other hand, a moderate increase of $\epsilon$ results in a more simple detection of the canard cycles. In particular, since the canard explosion occurs within an exponentially small range $\mathcal{O} (exp(-1/\epsilon))$ of the control parameter $K$, by increasing $\epsilon$, one can more easily approach the transition and find canard orbits with a more extended portion flowing close to the repelling part of the manifold.

\begin{figure}
\begin{center}
\includegraphics[width=0.6\textwidth]{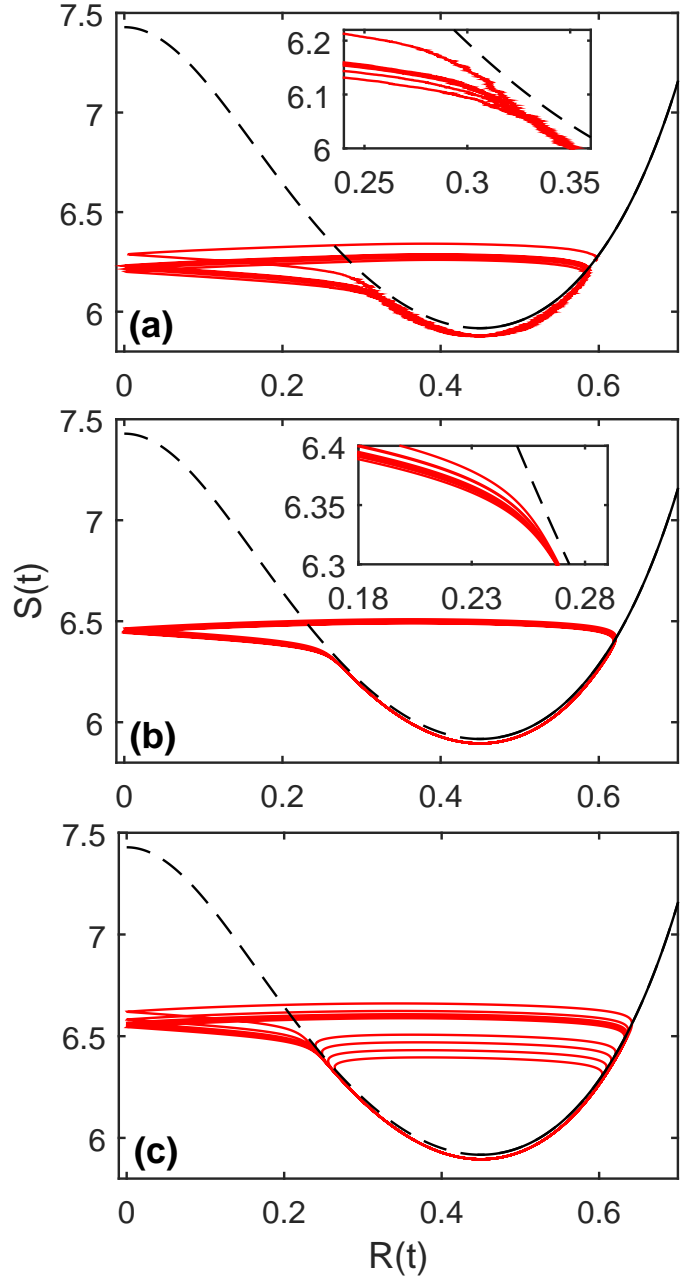}
\end{center}
\caption{Irregular canard cycles (red traces) for (a) the network model (\ref{network}) ($K=9.017134$) and (b,c) the mean-field model (\ref{mf}): (b) $K=$9.0444014469 (c) $K=$9.044401447113997. The cycles are illustrated in the ($S$,$R$) plane together with the critical manifold (\ref{smr}, whose attracting (repelling) branches are plotted as black solid (dashed) lines. Other parameters as in Fig. \ref{figure1}.}
\label{fig3}
\end{figure}

\begin{figure}
\begin{center}
\includegraphics[width=0.8\textwidth]{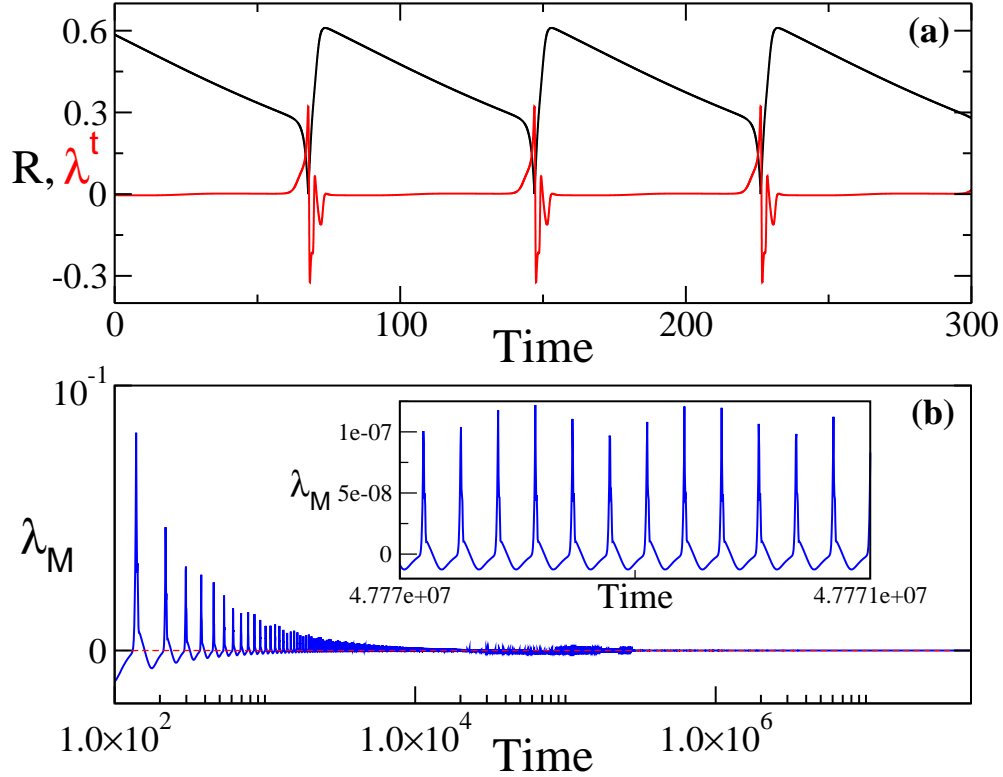}
\end{center}
\caption{(a) Time-dependent behavior of $R$ and $\lambda^t$. (b) Running average of the maximal Lyapunov exponent $\lambda_M$ vs time. In the inset is shown an enlargement of the time-dependent behavior for long simulation times.
Parameters: time step $\Delta t=0.001$, transient time $t_{tr}=2000$, simulation time $t_s=5 \times 10^7$, $K=9.044401447113997$, $\epsilon=0.02$. Other parameters as in Fig. \ref{figure1}.}
\label{fig.lyap}
\end{figure}

Further increasing $K$ towards the boundary of the Hopf bifurcation, such canard portion should continuously increase, eventually becoming maximal and thus following the whole repelling branch $\Sigma_R$. However, at difference with the planar systems in which canards are asymptotically stable, here we observe a transition from periodic to irregular canard cycles (see Fig. \ref{fig3} a)). The trajectories in the phase space display a maximal variability after their inflection point and generally, when they are far from the critical manifold. This is consistent with the fact that, close to the critical manifold, the dynamics is almost constrained on a plane. This behaviour has been reproduced by the mean-field model, demonstrating that the phenomenon persists in the thermodynamic limit and it is not related to finite-size effects. We found the irregular behaviour in the parameter range $K \in [9.04440144,9.0444014475]$. For higher values of $K$ in the interval, the irregular canard regime is a transient and the system eventually converges to a quasi-harmonic Hopf cycle (see Fig. \ref{fig3} c)). As $K$ is decreased starting from the upper bound, the time taken by the system to reach the Hopf cycles significantly grows. In Fig. \ref{fig3} b) we report the case for $K=9.0444014469$, where apparently stable irregular canards are observed. Further investigation on these irregular canards in terms of the Lyapunov exponent has clarified the origin of these irregularities. The behavior of the finite time Lyapunov exponent $\lambda^t$ is determined by the steep slopes shown by $R$ nearby the inflection point (see Fig. \ref{fig.lyap} a): when the order parameter rapidly leaves the critical unstable manifold, $\lambda^t$ is positive, but it suddenly becomes negative when $R$ is attracted back towards the stable manifold and it remains 0 along the stable manifold. The maximal Lyapunov exponent, whose running average is shown in Fig. \ref{fig.lyap} b) as a function of time, exhibits a clear tendency to reach $\lambda_M=0$ in the infinite time limit with a slope inversely proportional to the time. Thus confirming the overall periodic nature of the motion. The erratic behaviour seen in Fig. 3 (c) is due to the strong instability of the system in proximity of the repelling manifold, where $\lambda^t \simeq 0.3$,  which enhances infinitesimal integration errors. However, at variance with chaotic systems this perturbation is readily reabsorbed when the system moves along the stable manifold, characterized by a large negative $\lambda^t$.

\section{Emergent slow manifolds}
\label{emergent}

In the thermodynamic limit, canard explosions and burst oscillations in our network are explained in terms of the stability properties of a 1D critical manifold. Remarkably, such a geometric structure is not encoded in the single-node dynamics, but spontaneously emerges in the macroscopic behaviour of the network.

\begin{figure}
\begin{center}
\includegraphics[width=0.8\textwidth]{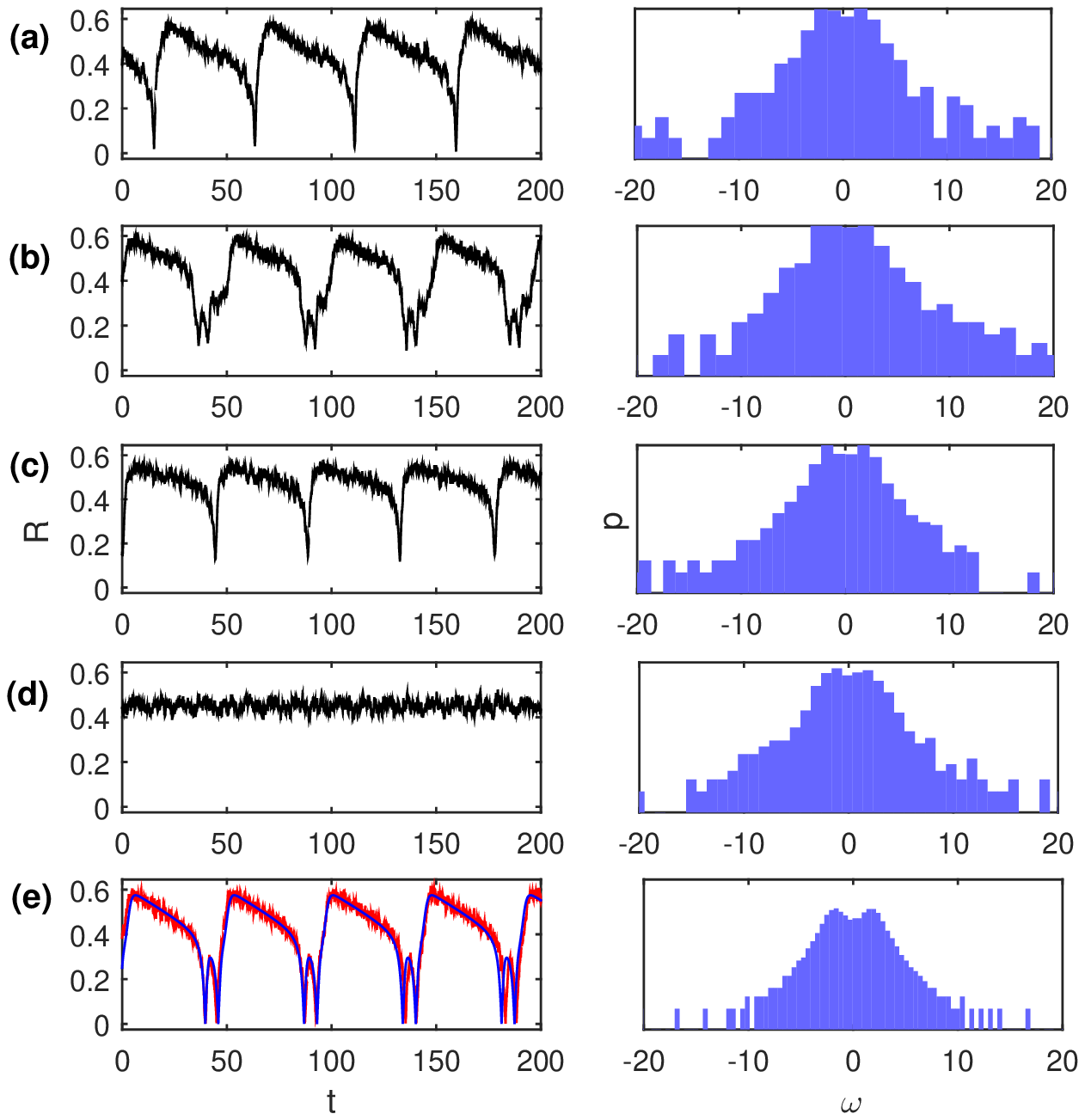}
\end{center}
\caption{Time-series of $R$ for the network model (\ref{network}) for $N=1000$ rotators and different sets of their natural frequencies. In (a-d) the frequencies are generated in a random way while in (e) are generated deterministically accordingly to the rule (\ref{gend}). In panel (e) we also show the corresponding mean-field prediction (blue solid trace). All parameters are fixed: $\epsilon=$0.02, $\Delta_0=$1.4 $\omega_0=$1.8, $\alpha$=7, $K=$8.955. The initial conditions are $\rho=$0.01,$\phi$=0, $S=$K.}
\label{figure4}
\end{figure}

To better understand such phenomenology we start investigating the macroscopic dynamics of the network for low values of $N$. In Fig. \ref{figure4}(a-d) we report the time-traces of $R(t)$ for $N=1000$ and different realizations of the frequency distribution. The results in (a-c) showing typical burst solutions clearly indicate that, already at low values of $N$, a collective slow-fast dynamics starts to emerge. On the other hand, when the distribution is randomly generated, the small modifications in the set of natural frequencies are sufficient to deeply affect the dynamics, with the system passing from a stable fixed point (d), to single (a,c) or multi-spike bursting (b). Such changes cannot be attributed to finite-size fluctuations, but they are rather the result of the dependence of the bifurcation points on each realization of the distribution. This variability indeed is neither observed when the distribution is generated deterministically through the rule (\ref{gend}), nor for different choices of the initial conditions when the distribution is randomly generated. In the latter case the population displays always the same macroscopic behaviour for the chosen parameters. We finally notice that, for a given $N$, the deterministic rule is more effective in approximating the mean-field dynamics. This is evidenced in Fig. \ref{figure4}(e) where, apart from the stronger finite-size fluctuations, the $2$-spike bursting is remarkably similar to that shown for a close parameter in Fig. \ref{figure1}(d). 

We now focus on collective canard cycles and analyze the dependence of the associated slow manifolds on the population size $N$. Since canard cycles contain portions of the attracting and repelling slow manifolds of the system, their study for different $N$ can help to visualize how effective manifolds for the macroscopic dynamics emerge. 

\begin{figure}
\begin{center}
\includegraphics[width=0.5\textwidth]{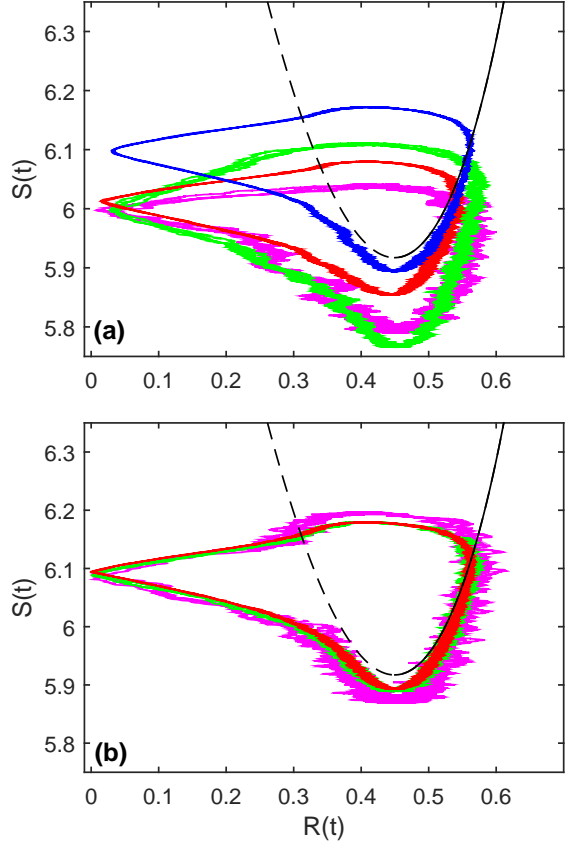}
\end{center}
\caption{Canard cycles in the ($S$,$R$) plane of the network model (\ref{network}) as a function of $N$ for (a) randomly-generated and (b) deterministically-generated Lorentzian frequency distributions. For each $N$ the control parameter $K$ has been adjusted to approach as much as possible the maximal canard. (a) $N=$1000, $K=$8.955 (magenta); $N=$5000, $K=$8.925 (green); $N=$20000, $K=$8.9470693 (red); $N=$50000, $K=$9.04 (blue). (b) $N=$1000, $K=$9.0246 (magenta); $N=$5000, $K=$9.035 (green); $N=$20000, $K=$9.038 (red). Other parameters as in Fig. \ref{figure1}.}
\label{fig5}
\end{figure}

In Fig. \ref{fig5}(a) we plot a number of canard solutions found for different network sizes. Owing to the dependence on the frequency distribution and sensitivity to fluctuations, the control parameter $K$ has been tuned to remain as close as possible to the transition between the Hopf cycle and the bursting regime, where canard trajectories are typically observed. As expected, the orbits display finite-size fluctuations that decrease as $N$ is increased. We observe that the typical features of the canard cycles are observable even for low values of $N$, although with a less evident inflection point. However, the trajectories seem to follow an effective slow manifold that clearly differ from $\Sigma$. A good agreement is recovered only for $N \gtrsim 20000$, while for lower values of $N$, the shape and location of the canard cycles in the phase space strongly depends on the realizations of the frequency distribution. This is suggestive of a direct dependence of the manifold geometry on the specific set of natural frequencies.
Such a dependence can be understood in terms of the following qualitative argument. The critical manifold of the network (\ref{network}) (i.e. in the singular limit $\epsilon=0$) is defined by the stable and unstable solution branches, corresponding to partially synchronized states of the fast dynamics ruled by Eq. (\ref{net1}), with $S$ acting as a constant coupling parameter. In the thermodynamic limit, these branches are determined by the "locked" rotators, labeled by the indexes $k$ such that $\vert \omega_k \vert \leq S R$. The contribution of the "drifting" oscillators with $\vert \omega_k \vert \geq S R$ instead vanishes in the case of symmetric frequency distributions \cite{strogatz}. For finite $N$ these branches, which perturb to effective slow manifolds for finite values of $\epsilon$, thus depend not only on the kind of frequency distribution (i.e. unimodal or bimodal), but also on the way in which the distribution has been generated. In particular, we expect a dependence on the relative weight between locked and drifting sub-populations, corresponding to the central part and the tails of the distribution respectively, the symmetry and the smoothness of the frequency distribution.

The above interpretation is supported by the results in Fig. \ref{fig5}(b), where we show canard orbits as a function of $N$ in the case of deterministically-generated frequency distributions. The deterministic rule allows to uniformly cover the range of possible frequencies leading, even for low $N$, to a comparatively smoother bimodal distribution. This leads to a reduction of finite-size effects and to a more rapid convergence to the mean-field dynamics \cite{nota}.
Already for $N=1000$, the canard cycle displays indeed the typical shape as observed in Fig. \ref{fig2}(a) with a considerable portion of the orbit lying in the vicinity of the repelling branch $\Sigma_R$. Even for such a low $N$, and in spite of the differences due to the finiteness of $\epsilon$, the 1D critical manifold of the mean-field model (\ref{smr}) provides a very good approximation of the effective slow manifold of the network. This is particularly evident in the vicinity of the attracting part (solid curve in Fig. \ref{fig5}) where the effects of the fluctuations and the sensitivity to the control parameter are lower.   

\section{Conclusions and Perspectives}
\label{concl}

We have shown that macroscopic canards cycles, special trajectories partially following invariant repelling slow manifolds, arise in the collective dynamics of a population of globally-coupled phase-rotators subject to adaptive coupling.
In particular, we considered the bimodal Kuramoto model, that it is known to display the coexistence of asynchronous and partially synchronized regimes, with an adaptive coupling. The coupling is controlled via a global linear feedback dependent on the level of synchronization of the network.

Canard solutions are related to locally invariant slow manifolds near non-hyperbolic points that, for sufficiently small $\epsilon$ (the ratio between the fast and the slow time-scale), are approximated by the set of equilibria of the fast dynamics. Remarkably, such a geometric structure is not encoded in the single-node equations, but is determined by the stable and unstable solution branches corresponding to partially synchronized states of the globally-coupled system Eq. (\ref{net1}) (with $S$ as a constant coupling parameter). We have studied collective canard cycles for different system sizes by considering both random or deterministically generated natural frequencies showing indeed that an effective slow manifold gradually emerges as the size of the population increases. Our results indicate that its geometry and the related canard orbits are determined not only by the kind of frequency distribution, but also by the uniformity of the set of natural frequencies.

Similar phenomena are likely to be observed also in different models, provided that the network dynamics without feedback display a hysteretic phase transition connecting a low synchronization state to one with a higher synchronization degree \cite{noi}. This is the case indeed of the Kuramoto model with inertia, in the presence of both unimodal or bimodal distributions. Our study thus opens interesting perspectives in these systems in which a low dimensional mean-field description has not been derived, for instance, for reconstructing the emergent slow invariant manifolds in the thermodynamic limit.  




\begin{thebibliography}{00}
\bibitem{nota} As an example of this convergence, we observed that similar accuracy with respect to the mean-field
results can be achieved by employing a network with $N=10,000$ rotators with the deterministic rule (\ref{gend}) and one with 
$N=1,000,000$ rotators with the randomly generated natural frequencies.
\end{thebibliography}


\end{document}